# Beating oscillation and Fano resonance in the laser assisted electron transmission through graphene δ- function magnetic barriers

by

R. Biswas[1] and C. Sinha[1,2]

[1]Department of Physics, P. K. College, Contai, Purba Medinipur, West Bengal- 721401, India

[2]Department of Theoretical Physics, Indian Association for the Cultivation of Science, Jadavpur-700032, India

Abstract: We investigate theoretically the transmission of electrons through a pair of δ-function magnetic barriers in graphene in presence of external monochromatic, linearly polarised and CW laser field. The transmission coefficients are calculated in the framework of non-perturbative Floquet theory using the transfer matrix method. It is noted that the usual Fabry-Perot oscillations in transmission through the graphene magnetic barriers with larger inter barrier separation takes the shape of beating oscillations in presence of the external laser field. The laser assisted transmission spectra are also found to exhibit the characteristic Fano resonances (FR) for smaller values of the inter barrier separation. The appearance of the perfect node in the beating oscillation and the asymmetric Fano line shape can be controlled by varying the intensity of the laser field. The above features could provide some useful and potential information about the light - matter interactions and may be utilized in the graphene based optoelectronic device applications.

**Introduction**:- During the last decade, the field of condensed matter physics involving graphene reigned the current research areas, particularly because of its huge amazing electronic, optical, thermal and chemical properties[1-3]. At the beginning, after the experimental realization of graphene[4,5], it was a great challenge for the scientist to create band gap in graphene, necessary for its exploitation in digital electronics. It is particularly due to the presence of the Klein Transmission (KT)[6,7], the fabrication of digital electronic circuitry along the line of conventional semiconductor technology could not be materialised by the use of graphene. One of the efficient methods to circumvent the problem was to use the external magnetic field[8-10] that is capable for breaking the dynamical symmetry of the system. The interaction with the substrate, the spin orbit coupling or the hydrogenation effect etc. might also be the other probable roots for this purpose[11].

The application of a uniform magnetic field on a graphene exhibits an extraordinary property of unconventional half integer quantum Hall effect[12] that could be explained by the existence of relativistic Landau Level formed by the charge carriers[13]. The interaction of an external uniform magnetic field with the chiral charge carriers plays an important role in manipulating the low energy properties of graphene which have been a subject of intense research in recent years[14]. On the other hand, the use of inhomogeneous magnetic field has given birth to the concept of magnetic barrier in a graphene with a view to confine the Dirac quasi-particles in the Landau levels that turns out to be an efficient tool to tailor the charge and spin transports (e.g., the suppression of the KT) in graphene based devices and has been studied exhaustively in last decade[8-10, 15-17]. The effect of an inhomogeneous magnetic field was already shown to exhibit interesting transport phenomena[18, 19], e.g., the magneto-resistance, the commensurable oscillation, the Shubnikov de Hass (SdH) oscillation, etc. in the study of the 2DEG conventional semiconductor hetero-structures. All these effects were also reported in the periodically modulated graphene in presence of the inhomogeneous magnetic field[20-23]. The periodic modulation in graphene can be considered either in the spatial domain (real space) or in the temporal domain (momentum space). The former can be realized through the application of time independent electrostatic or magnetic periodic potentials (also called graphene super-lattices) by the deposition of an array of parallel metallic or ferromagnetic strips on the surface[20,21,24-29]. On the other hand, the latter can be achieved by the use of a periodic time dependent potential or by an electromagnetic radiation (particularly the laser). Most of the earlier time dependent problems in graphene dealt either with the application of electromagnetic radiation on graphene[30-37] or with the application of a sinusoidal (AC) voltage on the bulk graphene[38] and graphene based quantum well/barrier structures[39-43]. It was pointed out by Trauzettel et. al.[38] that photon assisted electron transport is a direct probe for the energy dependent transmission in graphene and the study is relevant to observe the relativistic phenomenon like the Zitterbewegung in bench top experiments. A manifestation of relativistic phenomenon like the KT also persists even in presence of the time harmonic potential[42] similar to the case of the static barrier one. In case of oscillating quantum well or barrier in graphene, the characteristic Fano resonance (FR) was noted[40, 41] in the transmission spectra. The theoretical investigation of Gu et. al.[30] revealed that under the electric dipole approximation, when the electric field has significant coupling to the electron degrees of freedom, topological insulating properties could be induced in graphene.

The study of the effect of external laser field on the condense matter systems has gained a momentum due to the availability of high power, tuneable, linearly polarised laser and the free electron laser. The laser assisted electronic transport was studied for the quantum resonant tunnelling structures using conventional semiconductor heterostructures[44-47] as well as for the graphene[48-53]. It was already reported from the band structure calculations that the resonant interaction of the Dirac fermions in graphene with the external electromagnetic (EM) field leads to the formation of a dynamical gap between the conduction and the valence band in the quasi-particle energy spectrum that can be controlled by changing the intensity and the frequency of the EM field[32, 49]. This results to the suppression of the KT, since the field assisted Hamiltonian is non-commutative with the pseudo-spin operator leading to the violation of the pseudo-spin conservation for the process of transmission[52]. It is well known that the optical conductivity of graphene is very poor $\left[\sigma_0 = e^2/4\hbar\right]$ in the terahertz to far infrared regime leading to a severe limitation of the potential applications of graphene in the fields of electronics and photonics[36]. However, the situation could be highly improved by the application of a magnetic field, particularly for a single layer graphene (SLG) nano-ribbon where the conductance can be increased up to two orders of magnitude than that for SLG. Further, the study of periodic modulation potential (spatial) in presence of an external magnetic field perpendicular to the graphene plane has revealed[54] interesting physical effects on the transport response in the system. Although a very few works on the laser assisted quasi-particle transmission through graphene electrostatic barrier were reported, the use of the magnetic barriers under the laser field is still absent in the literature.

Further, the study of irradiated graphene microstructures bears the fundamental importance since the underlying quantum physics deals with the interaction between the two mass less particles – the photon and the Dirac fermion. The above discussions motivated us to investigate the simultaneous effect of the interaction of the chiral Dirac fermion with the laser and the magnetic fields by studying the laser assisted electronic transport through graphene magnetic barriers. Quantum bound states were already predicted[8, 10] for a magnetic step and for magnetic barriers of finite width, but none for δ- function barriers. However, it was reported that quasi-bound states exist between two consecutive δ- function magnetic barriers[55-57]. In this context, the present article aims at studying the effect of an external laser field on these quasi-bound states (between two delta function magnetic barriers with zero net magnetization) from the analysis of the Floquet transmission spectra obtained through the non perturbative approximation. Here we present the quantum interference effect that appears with precise finger prints, such as those produced by the Fano resonances.

The use of laser field provides us a number of controlling parameters in addition to those already existing for the magnetic barriers. It is worth mentioning that the control or manipulation of quasi-energy levels and transmission profiles in graphene based tunnelling structures is inevitable not only from the theoretical point of view but also for their successful exploitation in device fabrications, e.g., the study of the FR finds a wide range of practical situations particularly in sensing and switching applications[58].

Theoretical Model:

The effective Hamiltonian for a mass less Dirac fermion in presence of a magnetic field ($\vec{B} = \nabla X \vec{A}_b$) perpendicular to the plane of graphene monolayer is given by

$$H_0 = v_F \vec{\sigma}.[\vec{p} - e\vec{A}_b(x)] \qquad (1)$$

where $v_F$ is the Fermi velocity = $c/300$, '$c$' being the velocity of light; $\vec{\sigma} = (\sigma_x, \sigma_y)$ are the Pauli matrices representing pseudo-spin analogous to original spin; $\vec{p} = -i\hbar(\partial_x, \partial_y)$ are the momentum operator in the graphene plane ($x$, $y$); '$e$' being the electronic charge; $\vec{A}_b(x)$ being the vector potential that is uniform along the y-direction but varies along the $x$- direction.

In order to avoid the complexity of the problem (as a first attempt for tunnelling through laser assisted magnetic barrier in graphene), we consider a model system of two delta function magnetic barriers, identical in height but opposite in direction and separated by a distance '$L$' such that the total magnetic field vanishes across the structure. The corresponding vector potential profile polarized along the y-direction ($\hat{y}$) can be given by;

$$\vec{A}_b(x) = B\hat{y} \quad \text{(in units of } B_0 l_0\text{)} \quad \text{for } 0 < x < L$$

$= 0$ elsewhere. (2)

where $l_0 = \sqrt{\frac{\hbar}{eB_0}}$ is the length scale with a typical magnetic field strength $B_0$; '$L$' being the separation between the barriers (also termed as the width of the vector barrier) directed oppositely and perpendicular to the (x, y) plane and '$B$' being the height of the barrier. The above potential profile can be created by depositing a ferromagnetic strip on top of the graphene layer[57, 59].

The above Hamiltonian (Eqn. (1)) satisfying the equation of motion $H_0 \varphi(x, y, t) = E \varphi(x, y, t)$ suggests a two component wave function of the form $(x, y, t) = (\varphi_a(x, y), \varphi_b(x, y))^T exp(-iEt)$, $E$ being the energy of the particle. The component waves (denoted by the suffices '$a$' and '$b$') referring the charge carriers at the two lattice sites obey the following two coupled differential equations[52],

$$\left[-i\frac{\partial}{\partial x} - i\frac{\partial}{\partial y} + A_b(x)\right]\varphi_b(x, y) + E\varphi_a(x, y) = 0 \qquad 3(a)$$

$$\left[-i\frac{\partial}{\partial x} + i\frac{\partial}{\partial y} - A_b(x)\right]\varphi_a(x, y) + E\varphi_b(x, y) = 0 \qquad 3(b)$$

Let us now assume that the region ($0 < x < L$) between the two δ- function magnetic barriers is illuminated by a monochromatic (of frequency υ) laser field, linearly polarized in the y- direction. Under electric dipole approximation, the laser field is represented by the vector potential $\vec{A}_l(t) = (0, A_{ly}(t), 0)$ where the time dependence is given by $A_{ly}(t) = A_0\cos(\omega t)$. $A_0$ and ω are the amplitude and angular frequency of the laser field. The corresponding electric field is given by $\vec{E} = -\frac{\partial \vec{A}_l}{\partial t}$. For this approximation wavelength (λ) of laser must be greater than the region of interaction, i.e., $\gg L$. The left (region-I) and right (region-III) leads are taken adiabatically. Under this consideration, the charge carriers are injected and collected from the magnetic barrier system via two highly doped, ideal leads on both sides so that the effect of the laser field could be neglected in the regions I and III while the transport through the region II is quantum coherent[30].

In presence of the external laser field, the Hamiltonian for the region II ($0 < x < L$)

$$H = v_F \vec{\sigma}.[\vec{p} - e\vec{A}_b(x) - e\vec{A}_l(t)] \qquad (4)$$

satisfies the corresponding time dependent wave equation $H\Psi(x,y,t) = i\hbar \frac{\partial \Psi(x,y,t)}{\partial t}$. Considering the same form of time dependence of the electron in both the sub-lattices, the full wave function in region II is chosen to be of the form $\Psi(x,y,t) = (\Psi_a(x,y), \Psi_b(x,y))^T f(t) exp(-iEt)$.

To find $f(t)$ in the above expression, we follow the same iterative method as adopted in our earlier works[51,52]. Using the set of equations 3(a) and 3(b) we find $f(t) = e^{-i\alpha \sin \omega t} = \sum_{n=-\infty}^{\infty} J_n(\alpha) e^{-in\omega t}$ and then using this $f(t)$ one can find an infinite set of coupled differential equations to be satisfied by $\Psi_a(x,y)$ and $\Psi_b(x,y)$ as

$$\frac{d\Psi_b(x)}{dx} + (k_y + B + m\omega)\Psi_b(x) - i(E + m\omega)\Psi_a(x) = 0 \qquad 5(a)$$

$$\frac{d\Psi_a(x)}{dx} - (k_y + B + m\omega)\Psi_a(x) - i(E + m\omega)\Psi_b(x) = 0 \qquad 5(b)$$

where we have used the fact that the y-component of momentum ($k_y$) is conserved throughout the structure and the corresponding y- component of wave function is taken as $\sim e^{ik_y y}$.

Finally, solving these two coupled differential equations 5(a) and 5(b), one can find the functions $\Psi_a^m(x)$ and $\Psi_b^m(x)$ containing a new index $m$ corresponding to the Floquet side band energy states. Therefore, taking into account the effect of the side bands in presence of the laser field, the full solution for the pseudo spin state in the inter barrier region – II can be obtained as

$$\Psi(x,y,t) = \sum_{m,n} C_m \left( \frac{1}{q_x^m + i\{k_y + B + m\omega\}} \right) e^{iq_x^m + ik_y y} e^{-i(E+n\omega)t} J_{n-m}(\alpha)$$

$$+ \sum_{m,n} D_m \left( \frac{1}{-q_x^m + i\{k_y + B + m\omega\}} \right) e^{-iq_x^m + ik_y y} e^{-i(E+n\omega)t} J_{n-m}(\alpha)$$

where the argument of the Bessel function $J_{n-m}$ of order ($n$-$m$) is $\alpha = A/\omega = F/\omega^2$, $(q_x^m)^2 = (E+m\omega)^2 - (k_y + B + m\omega)^2$ and $C_m$, $D_m$ are the constant coefficients in the region II. The regions I and III are of the same form as in our previous works[51,52].

Finally, matching the pseudospin components at the two barriers ($x = 0$ and $x = L$), one can find the transmission coefficient ($T_m$) for the $m^{th}$ sideband of energy ($E + m\omega$) and the amplitude of transmission $E_m$ by using the relation[42,51]

$$T_m = \frac{\cos\theta_m}{\cos\theta_0} \left| \frac{E_m}{A_0} \right|^2 \qquad (7)$$

where $\theta_m = \tan^{-1}\left(\frac{k_y}{k_x^m}\right)$ and $(k_x^m)^2 = (E+m\omega)^2 - (k_y)^2$. Here $A_0$ is the amplitude of the wave incident with energy $E$ and angle $\theta_0$ in the region I.

**Results and discussions:**

To study the effect of irradiation on the tunnelling spectra of the Dirac fermion in graphene through the δ- function magnetic barriers, let us first recapitulate the transmission spectra in absence of the external laser field for $L = 50$, $B = 2$ and $k_y = 2$ as shown in Fig.2a. Above mentioned time independent problem was studied earlier by different groups of authors[55-57]. It is worth mentioning that the energy dependent transmission under the laser free condition is highly oscillatory in nature (due to the Fabry-Perot (FP) resonance) with the amplitude of oscillation damped monotonically while maintaining the perfect transmission

($T_c \sim 1$) at the resonance maxima. It is well known in optics that the FP oscillation arises due to multiple reflections of the light waves between two parallel plates. In analogy to this the present aforesaid oscillation verifies the wave nature of the Dirac fermions. The number of resonant peaks increases with the increase in barrier separation. On the other hand, with the increase in strength of the δ –function magnetic barrier, the magnitude of the $T_c$ decreases at the minima (i.e., the depth of the minima increases), showing the possibility of sharp confinement of the charge carriers in between the barriers.

In order to apprehend the effect of the external laser field, we display in Fig.2 (b) the $T_c$ for ω = 1 and *F* = 1 (other parameters same as Fig. 2(a)). Characteristic beating oscillations are noted in the energy dependent transmission profile in presence of the laser field. It is clear that the effect of the laser is to suppress the resonance transmission and the amount of suppression exhibits almost a periodic behaviour with respect to the energy of the electron incident on the barrier. As is well known for the sound wave, the formation of the beating oscillation, i.e., the creation of nodes and anti-nodes are due to the superposition of two waves with slightly different frequencies. The present beating pattern indicates that the electron transmission is strongly modified around the small energy windows corresponding to the nodal points. This could probably be explained as follows. The frequency of oscillation in the field free (FF) transmission for the two single δ- function barriers is supposed to be identical and as such the beat phenomenon that results from the interference of two sources having similar (but not identical) frequencies does not occur here under the FF condition. While, in presence of the external laser field, the single oscillation frequency for the two δ –function barriers gets modified creating a difference between the two individual frequencies ($f_1 \neq f_2$). This results in the occurrence of the quantum beats in presence of the external laser field due to the superposition of the two closely spaced frequency components with similar amplitudes. The frequency of the beats is expected to be equal to the difference between the two frequencies ($f_1$-$f_2$). The appearance of a perfect node depends actually on the intensity of the laser field as is evident from Fig.2(c). It is also interesting to note that the energy separation between two consecutive nodes increases with the increase in incident energy. This is probably due to the increase in the separation between the transmission resonances with the increase in energy, as noted from Fig. 2(a).

Figs. 3(a) and 3(b) show the variation of the transmission with respect to the change in the strength of the magnetic field respectively under the FF and laser assisted condition of the system. The FF (without laser) situation exhibits the so called Sd-H oscillation[23] whereas the results under the laser field display the beating nature of the Sd-H oscillation as explained in the earlier section.

The beauty of the Floquet approach lies in the fact that in this case one can find the probability of transmission in each of the photon assisted (emission or absorption) as well as in the no photon processes. Thus to study the nature of the transmission through different Floquet sidebands, we have displayed (Figs. 4(a-d)) the $T_c$ for the single photon absorption ($T_{+1}$), the single photon emission ($T_{-1}$) and for no photon ($T_0$) processes for the systems with fixed $L = 50$ and $k_y = 2.0$ but for $B = 2$ or 3, F = 1.0 or 1.1 and $\omega$ = 1.0 or 1.1. Fig. 4(a) reveals that the transmissions for all the aforesaid processes are oscillatory in nature and the mean transmissions (for all the incident energies) are greater for the central band (no photon exchange) than that for the other photon exchange processes. There is a clear competition of transmission among the different side bands while the average transmission profile in each side band shows resonant like peaks with larger half width and $T_c$ much less than unity as compared to the field free situation. Due to the irradiation by laser in the inter-barrier region, the incident particle flux is redistributed over the Floquet sidebands, although the sideband transmissions are less oscillatory than the central band. With the decrease in frequency of the laser field, the probability of the photon exchange transmission increases as compared to the no photon case (vide figs. 4(a) and 4(b)). On the other hand, Figs. 4(a) and 4(c) depict that the maxima (magnitude) of the side band transmission increase with the increase in the laser intensity. Further, with the increase in strength of the magnetic barrier, the amplitude of oscillation of the $T_c$ increases significantly particularly near the peaks and dips for the central band but for the other bands near the peaks only (vide Figs. 4(b) and 4(d)). Due to the increase in $B$, the energy separation between the adjacent peaks (of the transmission envelope) also increases in the central as well as in the sideband spectrum.

The effect of the barrier separation on the Floquet transmission is shown in Figs. 5(a) and 5(b) which represent the results ($T_0$, $T_{\pm 1}$) for $L = 10$ and $L = 2$ respectively. As we have already mentioned, the number of FP resonant peaks within a particular energy interval decreases with the decrease in the inter barrier separation $L$. Though the transmission remains oscillatory in nature, the variations of $T_0$ and $T_{\pm 1}$ are just the reverse for the energy above and below 6.5 for the case with $L = 10$. With further decrease in length ($L = 2$), the number of resonant peaks as well as the sharpness of the resonances of the field free transmission decrease as is noted from Figs. 5(a) and 5(b). The overall effect of the laser field (dotted line for FF and solid line with field, Fig. 5(b)) is to reduce the amplitude of oscillation, while retaining all other features (e.g., positions of crest and tough, average transmission etc.) unchanged. The intersections between the field free and the field assisted transmission indicate that for certain energies the laser has no effect on the tunnelling Dirac fermions through the magnetic barriers. So far as the individual band transmission is concerned, we find that the probability of transmission through the central band (zero photon exchange) decreases with the increase in energy, in sharp contrast to the increase in transmission for the photon exchange (absorption or emission) processes.

To study the effect of laser coupling parameter $\alpha = F/\omega$ on the kinetic transport of the charge carriers, we choose the parameters as $L = 2$, $B = 1$, $\omega = 1.1$, $\theta = 30^0$ and $E = 5$, the corresponding results are displayed in Figs. 6(a) and 6(b). It is clear from the figures that for lower intensity and high frequency of the laser field, the transmission occurs mainly through the central band and the higher energy Floquet sidebands are less accessible to the transmitted electrons as compared to the lower ones. The oscillatory nature of the transmission through different Floquet states follows from the nature of variation of the Bessel function with respect to its arguments, somewhat similar to the case of oscillating electrostatic barriers as mentioned earlier[42].

The reduction in the inter-barrier separation provides an interesting feature of the laser assisted tunnelling through the $\delta$ – function magnetic barriers as follows. The transmission profiles (as shown in Figs. 7 (a-c)) exhibit the characteristic Fano Resonance (FR), a phenomenon of quantum interference between the discrete and the continuum states that occurs here due to the interaction of the Dirac fermions with the oscillating field. In presence of the laser, the quasi-bound state between the delta-function magnetic barriers may provide the discrete channel of scattering required by the Fano resonance to occur as shown in Fig. 7(a-c). Thus the presence of the FR in the laser assisted transmission spectrum clearly dictates[39-41] the position of the quasi-bound states between the magnetic barriers. On the other hand, from the knowledge of the quasi-bound states, one can calculate the external laser frequency from the study of the Fano spectrum and thereby the tunneling structure might act as a radiation detector. Further, the FR has been demonstrated both theoretically and experimentally to be an important probe to reveal the properties of graphene[39].

With the decrease in laser intensity, the FR disappears indicating the cloaking effect of the magnetic barriers. It means that the low intensity laser field could not sense the presence of the quasi-bound states leading to the non-existence of the FR in the transmission spectra. From Fig.7(c) we can conclude that the FR for the central band as well as for the sidebands occur around the laser coupling parameter α = 3.47.

Conclusion: We have studied the magneto - radiative effects on the transport property of the Dirac fermions in a graphene based microstructure. The magnetic vector barrier is created using two δ- function magnetic fields of equal strength, applied perpendicular to the graphene sheet in opposite directions and separated by a distance '$L$'. Along the line of Floquet approach we calculate the transmission probabilities in the different Floquet side bands that arise due to the exchange of photons by the electron with external radiation field. In absence of the laser field the transmission through a single vector barrier exhibits Fabry-Perot like resonance ($(T_c)_{max} = 1$) where the number of peaks within a given energy range increases with the increase in '$L$' and the height of minima decrease with the increase in the strength of the magnetic field. By the application of the laser field the FP oscillation takes the shape of beating oscillation with the suppression of transmission at the resonance maxima. The effect of the laser field is expected to be maximum at the nodal points and is found to depend particularly on the intensity of the laser field. Such a beating oscillation may arise due to the commensurability effect of the periodic potential and the magnetic field in the graphene based structure. It is also noted that the accessibility of higher sidebands increases with the increase in intensity and the decrease in frequency of the laser field. Another interesting observation of the present study is the occurrence of the asymmetric Fano resonance that arises due to the quantum interference of the quasi-bound hole state inside the barrier with the electron continuum via the exchange of photon with the external field. The detection of such FR provides an efficient tool to identify the quasi-bound state inside the barrier not yet reported in case of the graphene magnetic barrier structure.

Figure Captions:

Fig.1: (a) Magnetic field profile corresponding to a pair of δ- function magnetic barriers of strength '$B$' and separated by a distance '$L$'. (b) Magnetic vector potential profile corresponding to the inhomogeneous magnetic field shown in (a).

Fig.2: Total side band transmission $T_c$ ( $\sum_m T_m$ ) plotted as a function of incident energy ($E$) for '$B$' = 2, $k_y$= 2 and '$L$' = 50. (a) Without laser field; (b) '$F$' = 1 and ω = 1; (c) '$F$' = 1.2 and ω = 1.

Fig.3: Total side band transmission $T_c$ ($\sum_m T_m$) plotted as a function of magnetic field ($B$) for '$E$' = 2, $k_y$= 2 and '$L$' = 50. (a) Without laser field; (b) '$F$' = 1.2 and $\omega$ = 1.

Fig.4: Transmission coefficients ($T_c$) for three individual side bands ($T_0$→ for no photon exchange, $T_{+1}$→for single photon absorption and $T_{-1}$→for single photon emission processes) plotted as a function of incident energy for $k_y$= 2 and '$L$' = 50. (a) for '$B$' = 2, '$F$' = 1.1 and $\omega$ = 1; (b) for '$B$' = 2, '$F$' = 1 and $\omega$ = 1; (d) for '$B$' = 2, '$F$' = 2 and $\omega$ = 1. (d) for '$B$' = 3, '$F$' = 1 and $\omega$ = 1.

Fig.5: Same as Fig.4 but '$F$' = 1, $\omega$ = 1 and $k_y$= 2. (a) for '$B$' = 3 and '$L$' = 10 ; (b) for '$B$' = 2 and '$L$' = 2 .

Fig.6: Individual band transmission plotted as a function of $\alpha$ for '$E$' = 5, '$B$' = 1, $\omega$ = 1.1 and '$L$' = 2. (a) $\theta_0 = -30^0$ and (b) $\theta_0 = 45^0$.

Fig.7: Same as Fig.4 but for '$L$' = 2. (a) for $k_y$= 1, '$B$' = 1, '$F$' = 5 and $\omega$ = 1.1; (b) for $\theta_0 = 30^0$, '$B$' = 1, '$F$' = 4.11 and $\omega$ = 1.1; (c) for $\theta_0 = 30^0$, '$B$' = 1, '$E$' = 5 and $\omega$ = 1.1.

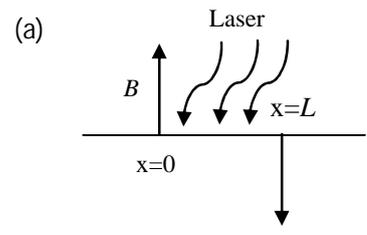

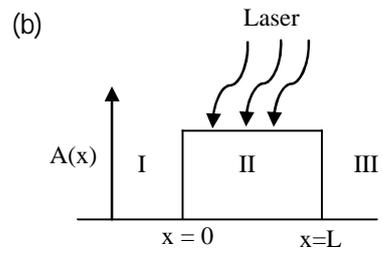

Fig. 1

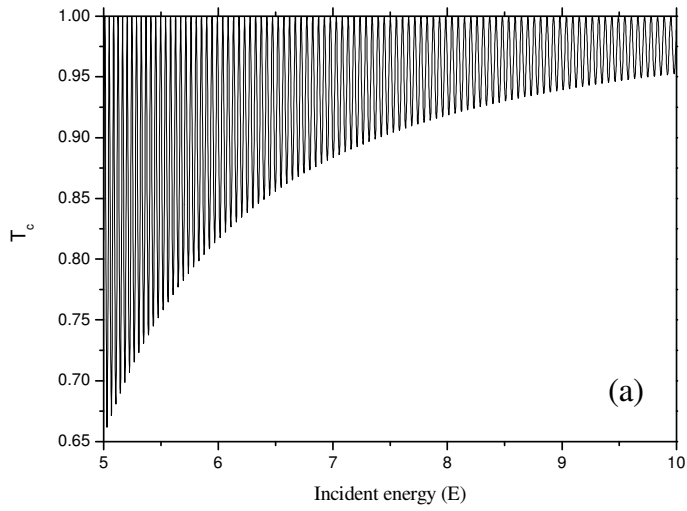

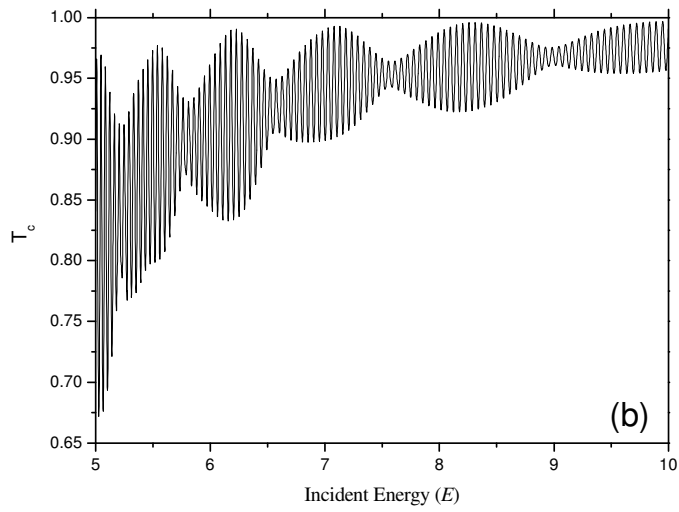

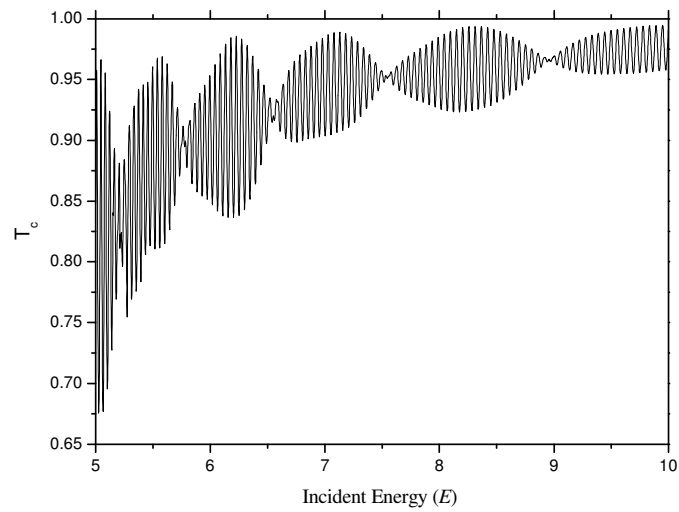

Fig.2

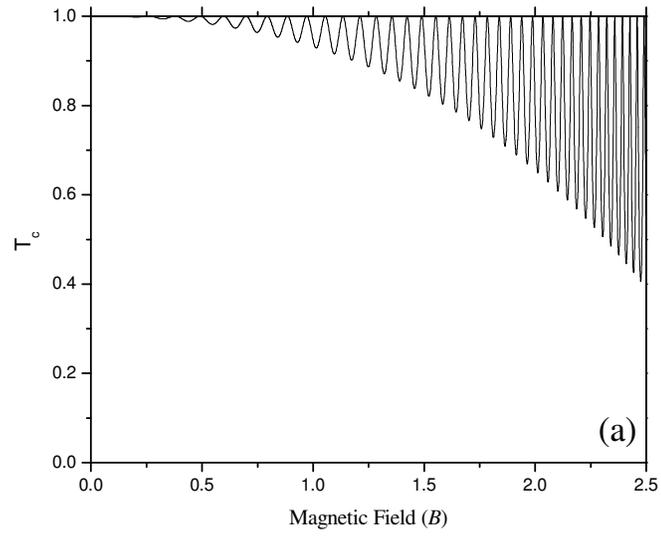

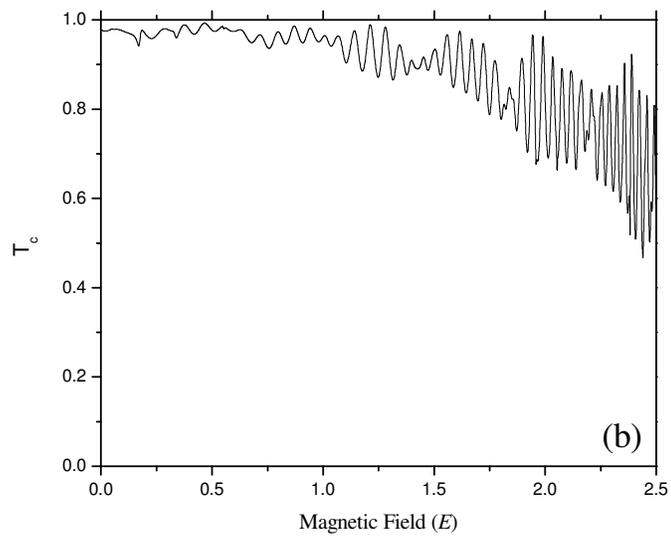

Fig.3

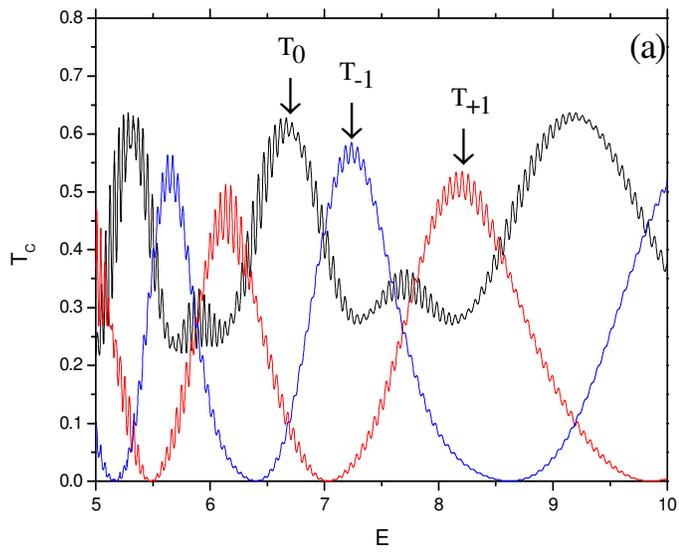
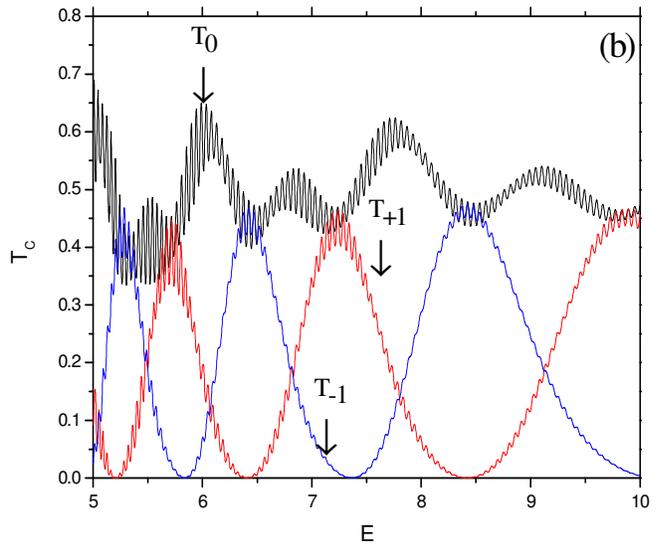
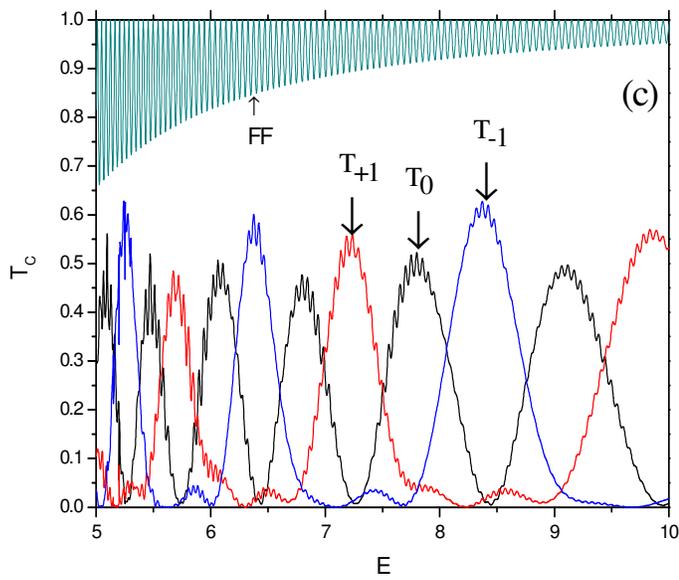
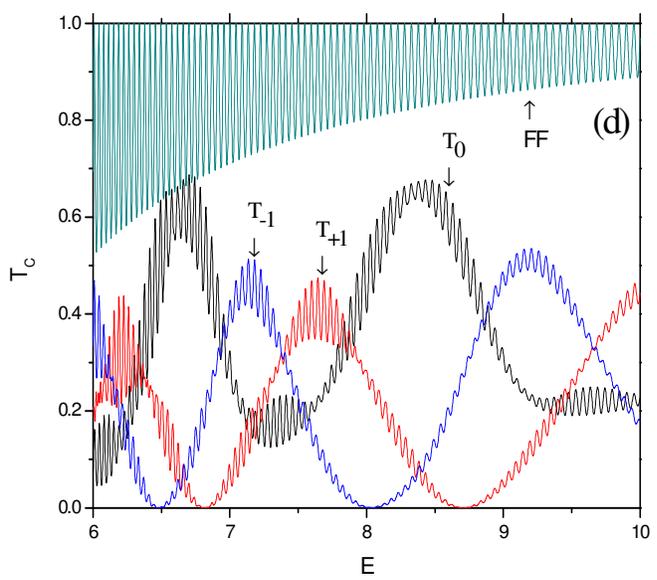

Fig.4

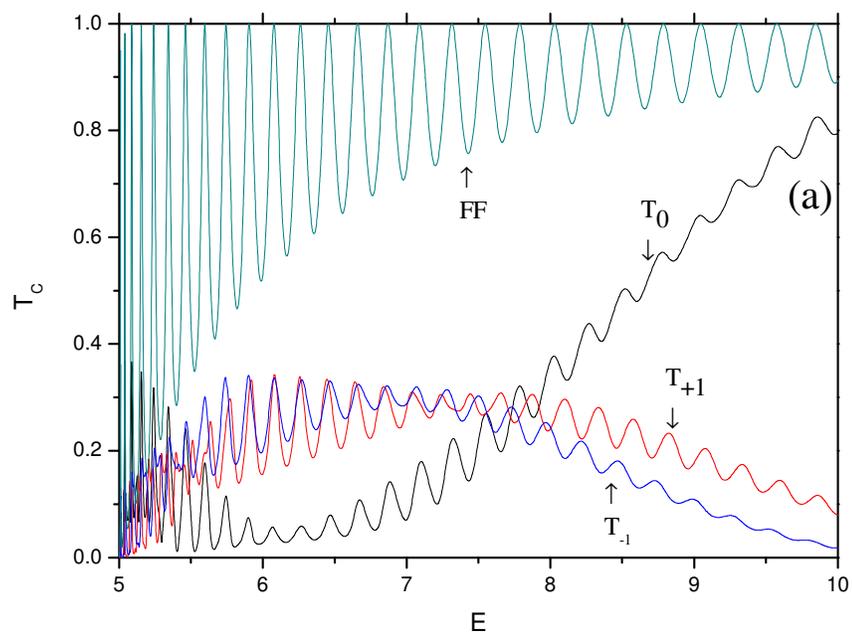

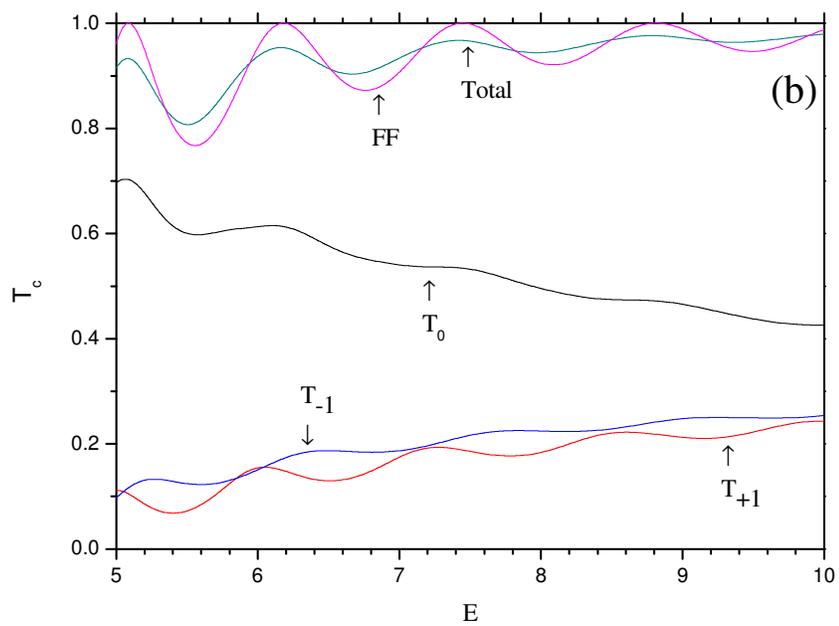

Fig.5

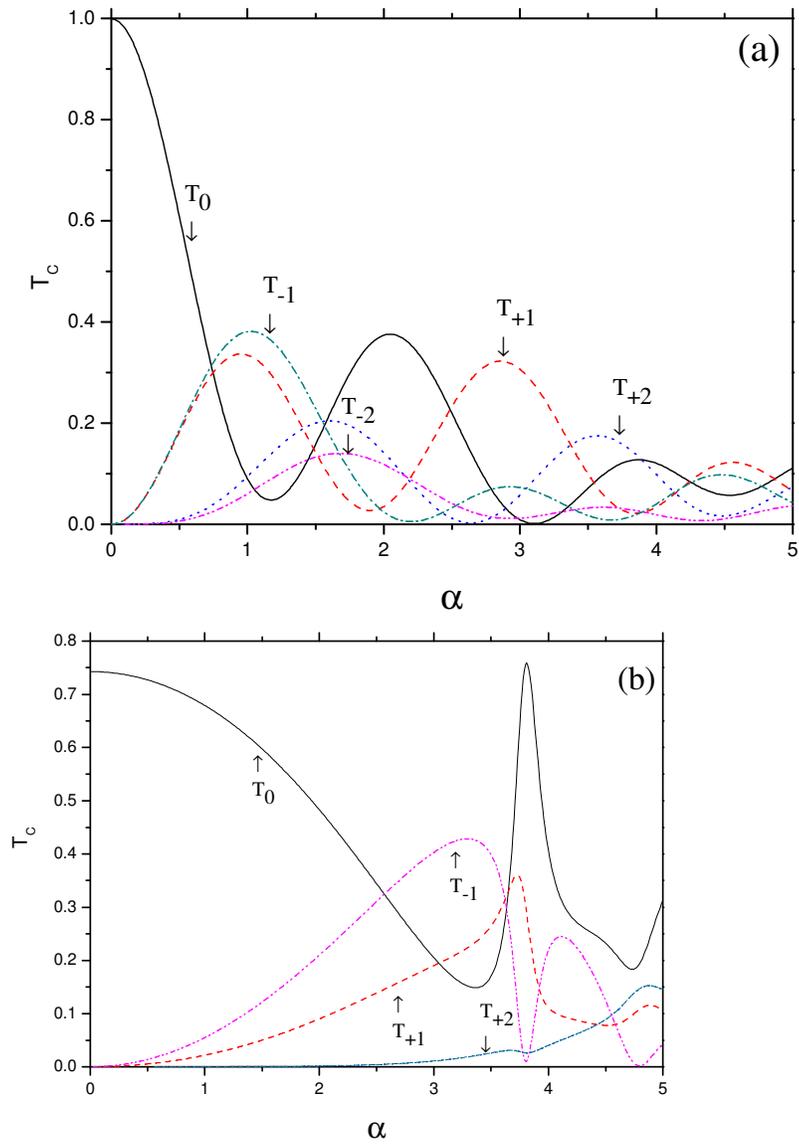

Fig.6

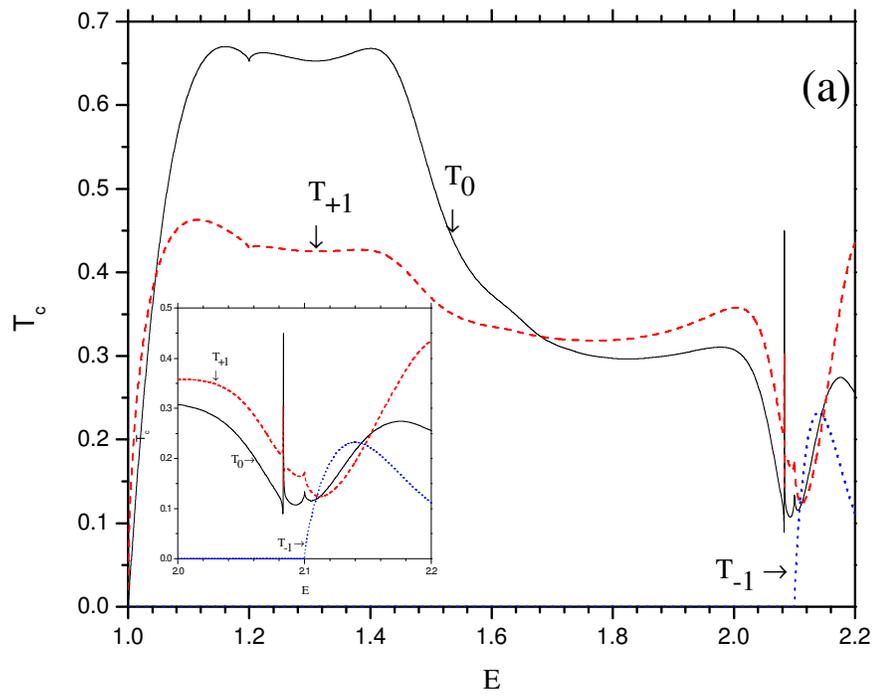
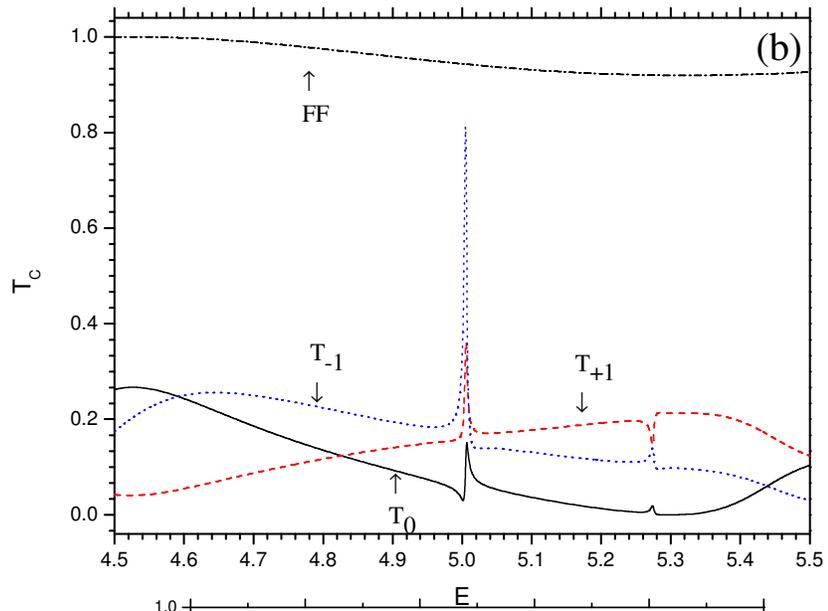
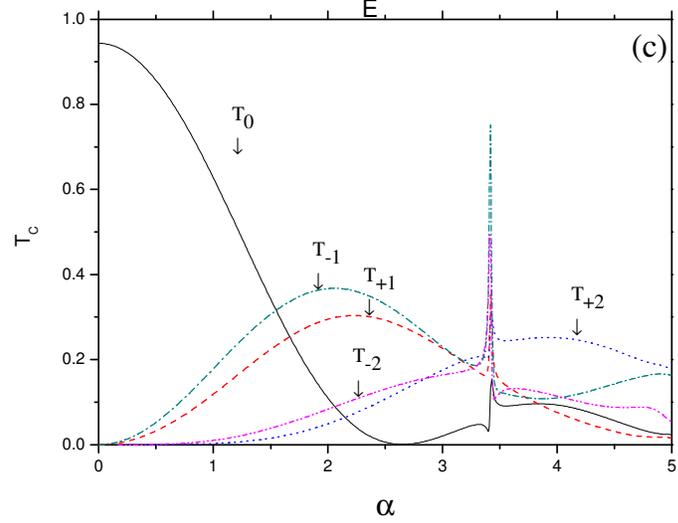

Fig.7